\documentclass[notitlepage,amsmath,superscriptaddress,amssymb,11pt,floatfix,aps,prd,nofootinbib,preprintnumbers]{revtex4-2}

\usepackage[margin=1in]{geometry}
\usepackage{setspace}
\usepackage{braket}
\usepackage{fancyhdr}
\usepackage{graphicx,wrapfig,lipsum}
\usepackage{float}
\usepackage{color,xcolor}
\usepackage{hyperref}
\usepackage{mathtools,amssymb,amsfonts,multirow}
\usepackage{bbm,dsfont}
\usepackage{array}
\usepackage{revsymb}
\usepackage{coffeestains}
\usepackage{bbold}
\usepackage{tikz-cd,pgfplots}
\usetikzlibrary{cd}
\usepackage{orcidlink}
\PassOptionsToPackage{unicode}{hyperref}
\PassOptionsToPackage{naturalnames}{hyperref}

\hypersetup{
	colorlinks=true,
    linkcolor=blue, 
    linktoc=all 
}

\newcommand{\de}{\mathrm{d}}
\newcommand{\del}{\partial}

\newcommand{\munu}{\mu\nu}
\newcommand{\ab}{\alpha\beta}

\newcommand{\und}{\quad\text{and}\quad}

\newcommand{\veq}{\mathrel{\rotatebox{90}{$=$}}}

\newcommand{\op}[1]{\operatorname{#1}}

\newcommand{\tr}{\op{tr}}

\renewcommand{\sf}[2]{\mbox{\small{$ \frac{#1}{#2} $}}}

\newcommand{\gfrak}{\mathfrak{g}}

\newcommand{\Acal}{\mathcal{A}}
\newcommand{\Fcal}{\mathcal{F}}

\newcommand{\SU}{\mathrm{SU}}

\newcommand{\GR}{\mathrm{GR}}

\makeatletter
\newcommand*\owedge{\mathpalette\@owedge\relax}
\newcommand*\@owedge[1]{%
  \mathbin{%
    \ooalign{%
      $#1\m@th\bigcirc$\cr
      \hidewidth$#1\m@th\wedge$\hidewidth\cr
    }%
  }%
}
\makeatother

\PassOptionsToPackage{unicode}{hyperref}
\PassOptionsToPackage{naturalnames}{hyperref}

\begin{document}


${}$\vskip1cm

\title{\textbf{\Large{Symmetric Yang--Mills theory in FLRW universes}}}

\author{Mahir~Ertürk\orcidlink{0009-0006-6205-6685}}
\affiliation{\mbox{Institut für Theoretische Physik, Leibniz Universität Hannover, Appelstraße 2, 30167 Hannover, Germany}}

\author{Gabriel Picanço\orcidlink{0000-0002-2698-607X}}
\email{G.Picanco@sussex.ac.uk}
\affiliation{\mbox{Institut für Theoretische Physik, Leibniz Universität Hannover, Appelstraße 2, 30167 Hannover, Germany}}
\affiliation{\mbox{Department of Physics and Astronomy, U Sussex, Brighton, BN1 9QH, U.K.}}
\affiliation{\mbox{Department of Physics, TU Dortmund, Otto-Hahn-Stra\ss e 4, 44227 Dortmund, Germany}}

\begin{abstract}\vspace{1cm}
\noindent\normalsize
In this work, we set up the theoretical framework and indicate future applications of symmetric Yang--Mills fields to cosmology. We analyze the coset space dimensional reduction scheme to construct pure Yang--Mills fields on spacetimes given as cylinders over cosets. Particular cases of foliations using $H^n$, dS$_n$ and AdS$_n$ slices as non-compact symmetric spaces are solved, compared to previous results in the literature, and generalized in a structured fashion. Coupling to general relativity in FLRW-type universes is introduced via the cosmological scale factor. For the hyperbolic slicing in 4D, the dynamics of the Einstein--Yang--Mills system is analytically solved and discussed. Finally, we generalize the analysis to warped foliations of the cylinders, which enlarge the range of possible spacetimes while also introducing a Hubble friction-like term in the equation of motion for the Yang--Mills field.
\end{abstract}

\maketitle

\newpage
\begin{spacing}{.9}
\tableofcontents
\end{spacing}


\section{\bf Introduction and summary}

The Yang--Mills theory was shown to be essential in many different areas in physics, most notably in the successful quantum description of the weak and strong interactions and the construction of the standard model of particle physics. Additionally, classical solutions of the Yang--Mills equation are key for other physical phenomena, as QCD confinement models \cite{SKYRME1962556}, the description of spin-orbit interactions in condensed matter physics \cite{Berche_2013}, and, most important for this work, gauge-flation scenarios in cosmology \cite{Maleknejad2012:1212.2921v3}.

Despite its relevance, the Yang--Mills equation, as a set of non-linear partial differential equations, is in general rather difficult to solve analytically. Therefore, methods allowing the construction of explicit analytic solutions may play a crucial role in obtaining a better understanding of the properties of non-Abelian gauge theories. One common approach for that is to impose additional symmetry conditions on the gauge field, oftentimes related to the structure of the underlying spacetime at hand, thus reducing the degrees of freedom of the Yang--Mills system. In particular, if the underlying spacetime can be described using a Lie group or cosets thereof, this additional structure may be used to further simplify the system of equations. In parallel with the establishment of the formal treatment of spacetime-symmetric Yang--Mills fields \cite{Bergmann:1978fi,Forgacs:1979zs,Harnad:1980,Molelekoa:1985if}, first applications in cosmology started to be investigated \cite{Henneaux:1982vs,HOSOTANI198444,Galtsov:1991un}, which later led to models of inflation driven by classical symmetric gauge fields (see \cite{Maleknejad2012:1212.2921v3} for a review).

The relation between the description of spacetimes through Lie group manifolds and classical solutions of Maxwell's or Yang--Mills' equations have recently found new applications. In \cite{EM_Knots}, the four-dimensional de Sitter space dS$_4$ was foliated with spacelike $S^3 \cong$ SU$(2)$ slices to obtain, through the conformal invariance of 4D Yang--Mills theory, an infinite basis of knotted electromagnetic solutions on Minkowski space, reproducing and generalizing the celebrated Hopf--Rañada solution \cite{Ranada}. Moreover, if the spacetime is a cylinder over a Lie group $M\cong\mathbbm{R}\times G$, one can impose invariance of the gauge field under the natural $G$-action on $M$. Then, the Yang--Mills equation reduces to a system of ordinary, albeit still non-linear, matrix differential equations, possibly with the number of degrees of freedom reducing to one. Multiple instances of such approach were used, for example, to construct analytic solutions on four-dimensional (anti-)de Sitter space (and hence by conformal invariance on Minkowski space) in \cite{dS4,Finite_Action,Exact}. Especially the case $M \cong \mathbbm{R}\times$SU$(2)$ saw applications related to cosmology and the electroweak epoch, where the Yang--Mills field was coupled to the scale factor of Friedmann--Lemaître--Robertson--Walker-type (FLRW) closed geometry \cite{friedan2020origin,Kumar_2021,DM,CGF}. A related approach to obtain symmetric gauge fields is the coset space dimensional reduction (CSDR) scheme \cite{Kapetanakis:1992hf}, where $G$-invariant Yang--Mills fields are considered on a spacetime which is foliated with cosets of $G$, that is, $M\cong \mathbbm{R}\times G/H$, where $H$ is some Lie subgroup of $G$. Recent applications of this approach include, for example, \cite{G2,Flows,Lechtenfeld_2018,Mink}. Especially in \cite{Lechtenfeld_2018}, the $(n{+}1)$-dimensional de Sitter space dS$_{n+1}$ was foliated using warped cylinders over $S^n$ written as cosets of the orthogonal, unitary, and spin groups; the first one, $S^n \cong $ SO$(n+1)$/SO$(n)$, also being a symmetric space. Moreover, non-compact cosets, and hence structure groups, made an appearance in \cite{Kumar_2021} to construct SO$(1,3)$ invariant solutions on Minkowski space by gluing solutions together, each part obtained on cylinders where different subgroups were modded out of the Lorentz group.

In this work, we consider the CSDR scheme applied both to straight $\mathbbm{R}\times G/H$ as well as warped $\mathbbm{R}\times_{a} G/H$ cylinders over (non-compact) symmetric spaces, where $a$ is a warping function. In particular, we consider pure Yang--Mills theory with gauge group $G$ with the cosets $G/H$ being hyperbolic space $H^n\cong$ SO$(1,n)$/SO$(n)$, de Sitter space dS$_n \cong$ SO$(1,n)$/SO$(1,n-1)$, and anti-de Sitter space AdS$_n \cong$ SO$(2,n-1)$/SO$(1,n-1)$. The general treatment and geometrical construction for straight cylinders is developed at the beginning of Section \ref{sec2:YM-CSDR}, where it becomes clear that the dynamics of the gauge field reduce to that of a one-dimensional Newtonian particle subject to a quartic potential. After that, we derive the reduced Lagrangians and equations of motion for the three cases at hand. We naturally find that the complementary scenario of $S^n \cong$ SO$(n+1)$/SO$(n)$ discussed in \cite{Lechtenfeld_2018} fits neatly into the results obtained for the three non-compact cosets, which we will from there on out be able to include into all later considerations. Subsequently, we compute a closed expression for the energy-momentum tensor of these scenarios, revealing their perfect-fluid structure for the two Riemannian slicings at hand. Finally, we give the analytic solutions to the equations of motion for the three systems, as they reduce to the dynamics of a Newtonian particle moving in an inverted or non-inverted double well. In Section \ref{sec3:GR}, we couple the $\mathbbm{R}\times H^3 $ case to gravity by considering an FLRW-type open hyperbolic cosmology. This is in analogy with the previous considerations of the $\mathbbm{R}\times \SU(2)$ case mentioned above. Naturally, the conformal invariance of the Yang--Mills theory in four dimensions simplifies the Einstein--Yang--Mills system to a one-way coupling which amounts to the Wheeler--DeWitt constraint, allowing us to solve the system analytically. In Section \ref{sec4:warping}, we generalize our CSDR setup to warped cylinders $\mathbbm{R}\times_{a} G/H$. We show that the introduction of such a warping results in the addition of a Hubble-friction term in the equation of motion for the analog particle. A hyperbolic slicing of AdS$_{n+1}$ is briefly discussed and compared to the aforementioned spherical slicing of dS$_{n+1}$ as a simple example of the developments above derived. In Sec.~\ref{sec5:conclusion}, we summarize our findings and indicate possible applications of the general framework here developed to cosmology.

\section{\bf Coset space dimensional reduction of Yang--Mills fields}\label{sec2:YM-CSDR}

In this section, we will recall some geometrical aspects that are relevant for this work and we will establish notations and conventions to be used.

Let $M$ be a (pseudo-)Riemannian manifold, $G$ a Lie group, and $\Acal \in \Omega^1(M,\gfrak)$ the local representative of a $G$ gauge theory over $M$, where $\gfrak = \mathrm{Lie}(G)$. If $\zeta$ is the generator of a diffeomorphism transformation in $M$, the condition
\begin{equation}\label{eq:Lie-derivative-of-A}
	\pounds_\zeta \Acal = 0\,,
\end{equation}
where $\pounds_\zeta$ is the Lie derivative with respect to $\zeta$, would be necessary and sufficient to impose invariance on a non-gauged field $\Acal$ with respect to the diffeomorphism. However, when $\Acal$ is a gauge field, such a requirement is not necessary anymore, since a non-trivial spacetime transformation of $\Acal$ may still be compensated by a gauge transformation without any change to the underlying physics. For a more comprehensive and pedagogical discussion of spacetime symmetries on Yang--Mills fields, we refer the reader to \cite{Forgacs:1979zs}, and, for a more precise formal definition, to \cite{Molelekoa:1985if}. That said, \eqref{eq:Lie-derivative-of-A} is still a sufficient condition and will be used in this paper.

In what follows, we will work with a pure Yang--Mills theory on a principal $G$ bundle $P=G\times M$ over a cylinder $M=\mathbbm{R}\times G/H$, where $G/H$ will be a symmetric space. In this section, the signature of the metric is not fixed. We will fix it later to mostly plus when we consider the spacetime dynamics. Let us now review the geometrical setup and then construct the coset space dimensional reduction of the gauge field.

\subsection{Spacetime geometrical structure}\label{subsec:geometry}

Given a Lie group $G$ and a closed (Lie-)subgroup $H\subset G$, we define the coset space $\tilde{M}:=G/H$. Let $\mathfrak{g}$ and $\mathfrak{h}\subset\mathfrak{g}$ be the Lie algebras of $G$ and $H$, respectively. We denote the Killing form on $\mathfrak{g}$ as
\begin{align}
K(X,Y):=\varepsilon_K \,\mathrm{tr}(\mathrm{ad}_{X}\circ\mathrm{ad}_{Y})\equiv\varepsilon_K \mathrm{tr}_{\mathrm{adj}}(XY)\;,\;\; X,Y\in\mathfrak{g}\,,
\end{align}
where we have introduced a factor $\varepsilon_K =\pm 1$ to adjust the overall sign of $K$. It can be used to remove a sign ambiguity which arises when we work with semi-simple groups $G$, that is, with indefinite Killing forms. As we will consider symmetric spaces, and thus reductive ones, the Lie algebra $\mathfrak{g}$ can be decomposed into Killing-orthogonal parts
\begin{align}
\mathfrak{g}=\mathfrak{h}\oplus\mathfrak{m}\
\end{align} 
such that
\begin{align}
[\mathfrak{h},\mathfrak{h}]\subset\mathfrak{h}\;,\;\;
[\mathfrak{h},\mathfrak{m}]\subset\mathfrak{m}\;,\;\; \text{and} \;\;
[\mathfrak{m},\mathfrak{m}]\subset\mathfrak{h}\,.
\end{align}
Let $\{I_A \}=\{I_\alpha \}\cup\{I_a\}$ be an orthogonal basis of $\mathfrak{g}$ with $\mathrm{span}\{I_\alpha \}=\mathfrak{h}$ and $\mathrm{span}\{I_a \}=\mathfrak{m}$. We choose the bases such that
\begin{align}\label{killing}
K(I_A ,I_B)=\varepsilon_K \,\mathcal{D}_{n}\,\tilde{\eta}_{AB}\,,
\end{align}
where $\mathcal{D}_{n}$ is a normalization depending on the dimension $n:=\mathrm{dim}(\mathfrak{m})$ of the coset, and where 
\begin{align}
\tilde{\eta}_{AB}:=\frac{\mathrm{tr}_{\mathrm{adj}}(I_A I_B)}{|\mathrm{tr}_{\mathrm{adj}}(I_A I_B)|} =\tilde{\eta}_{\ab}\oplus\tilde{\eta}_{ab}
\end{align}
is the normalized diagonal version of $\mathrm{tr}_{\mathrm{adj}}(I_A I_B)$. Now let $\{\hat{e}^A\}=\{\hat{e}^\alpha\}\cup\{\hat{e}^a\}\subset\Omega^1(G)$ be the corresponding basis of left-invariant one-forms on $G$.\footnote{That is, the dual basis to the left invariant vector fields $\hat{E}_A$ generated by left translation of the $I_A$.} With these, we get the Cartan--Killing metric on $G$ via
\begin{align}
g^{(\mathrm{CK})}=\tilde{\eta}_{AB}\,\hat{e}^A \otimes \hat{e}^B\,.
\end{align}
Turning our attention to the coset space, let $\sigma : G/H \rightarrow G$ be a (possibly local) section and consider the pullbacks $e^A := \sigma^{\ast}\hat{e}^A \in\Omega^1 (G/H)$. Since $\mathrm{dim}(G/H)=\mathrm{dim}(G)-\mathrm{dim}(H)=\mathrm{dim}(\mathfrak{m})$, the set of one-forms $\{e^A\}$ will be linearly dependent, with the subset $\{e^a\}$ forming a basis on the coset and with 
\begin{equation}
	e^\alpha =\chi^{\alpha}_{\;b}e^b\,,
\end{equation} 
for some functions $\chi^{\alpha}_{\;b}\in\Omega^{0}(G/H)$. Choosing the section appropriately, we obtain the `canonical' left-invariant (pseudo-)Riemannian metric on the coset via
\begin{align}
g_{\tilde{M}}=\tilde{\eta}_{ab}\, e^a \otimes e^b\,.
\end{align}
We remark that the orthonormality of the forms on the coset imply the first Cartan structure equation
\begin{align}
\de e^a + \omega^{a}_{\;\;b}\wedge e^b =0\,,
\end{align}
where $\omega^{a}_{\;\;b}$ is the Levi--Civita connection form, whilst the left invariance of the forms on $G$ implies the Maurer--Cartan equations both on $G$ and on the coset,
\begin{align}
\de e^A =-\frac{1}{2}f_{BC}^{\;\;\;A}\,e^B \wedge e^C\,.
\end{align}
Due to the fact that $G/H$ is symmetric, the equation above decomposes into two sets of independent equations. Both of these properties together can be used to express the Levi--Civita connection $\omega^{a}_{\;\;b}$ on $G/H$, and thus all related curvature quantities such as the Riemann tensor, the Ricci tensor, and the Ricci scalar purely in terms of the structure constants $f^{\;\;\;A}_{BC}$ and of the functions $\chi^{\alpha}_{\;b}$.

With the geometric structure of the coset laid down, we now define the Lorentzian spacetime as the flat cylinder $M:=\mathbbm{R}\times G/H$. Warped cylinders will be considered in Sec.~\ref{sec4:warping}. The metric on $M$ will be the product metric
\begin{align}\label{metric}
g:=\varepsilon_g\, \tilde{g}_{\munu}\,e^\mu \otimes e^\nu =\varepsilon_g \,(\tilde{g}_{00}\,e^0 \otimes e^0 + \tilde{\eta}_{ab}\,e^a \otimes e^b )\,,
\end{align}
where $e^0 \equiv e^u :=\de u$, with $u\in\mathbbm{R}$, is the foliation parameter and $\tilde{g}_{00}=\pm 1$ is chosen depending on the signature of $\tilde{\eta}_{ab}$. If the coset metric $\tilde{\eta}_{ab}$ is Lorentzian (Riemannian), the foliation parameter will be spacelike (timelike). We again introduce a factor $\varepsilon_g = \pm 1$ to keep track of the overall sign of $g$.

\subsection{Coset space dimensional reduction}

With the geometric structure of the spacetime $M$ laid down, we now consider the principal $G$ bundle $P=G\times M$ to study a gauge theory over $M$. We expand both the gauge field $\mathcal{A}\in\Omega^{1}(M,\mathfrak{g})$ and its field strength $\mathcal{F}=\de\mathcal{A}+\mathcal{A}\wedge\mathcal{A}\in\Omega^{2}(M,\mathfrak{g})$ in terms of the basis of one-forms defined above such that
\begin{align}
\mathcal{A}= \mathcal{A}_a\,e^a \;\; \text{and} \;\;\mathcal{F}=\mathcal{F}_{0a}\,e^0\wedge e^a +\tfrac{1}{2}\mathcal{F}_{ab}\,e^a\wedge e^b\,,
\end{align} 
where we chose to work on the `temporal' gauge $\mathcal{A}_0 \equiv 0$, which we will keep throughout this work. In general, the $\Acal_a$ coefficients could be any elements in the Lie algebra $\gfrak$, however, imposing left $G$-invariance on the gauge field $\Acal$ yields \cite{Kapetanakis:1992hf}
\begin{align}
\mathcal{A}=I_\alpha \, e^\alpha + \underbrace{X_{a}(u)}_{\in\mathfrak{m}}\, e^a\,,
\end{align}
where the $X_a$ are constrained by
\begin{align}
[I_\alpha ,X_{a}(u)]=f_{\alpha a}^{\;\;\;b}X_{b}(u)\,,
\end{align}
that is, the components $X_a(u)$ lie in the adjoint representation $\mathrm{ad}(\mathfrak{h})|_{\mathfrak{m}}$ of the subalgebra on the orthogonal complement\footnote{More concisely, the $X_{a}(u)$ must transform like the basis elements $\{I_a\}$.}. The constraint is solved by introducing a `scalar' degree of freedom $\phi_{i}(u)$ for each irreducible representation $\mathcal{R}_{i}$ of $\mathrm{ad}(\mathfrak{h})|_{\mathfrak{m}}=\mathcal{R}=\bigoplus_{i}\mathcal{R}_{i}$. This is accomplished by first changing the basis $\{I_a \}$ into new sub-bases $\bigoplus_i \{\bar{I}_{a_i}\}$, which block-diagonalize the representations, then scaling each sub-basis with $\phi_{i}(u)$ and finally transforming back to obtain the $X_a(u)$. Notice that this in general implies that the $X_{a}(u)$ mix generators from different irreps (see \textit{e.g.} \cite{Lechtenfeld_2018,G2}). However, in all cases considered in this work, the adjoint representation $\mathrm{ad}(\mathfrak{h})|_{\mathfrak{m}}$ is always the vector representation of $\mathfrak{h}$ and thus irreducible. Hence, we reduce the system down to a single degree of freedom $\phi(u)$, allowing us to set $X_{a}(u)=\phi(u)\,I_a$. With that, the gauge field is decomposed simply as
\begin{align}\label{eq:A-reduced}
\mathcal{A}=I_\alpha e^\alpha +\phi(u)\,I_{a}\,e^{a}\, ,
\end{align}
from which we get the components of the field strength
\begin{align}\label{ansatz}
\mathcal{F}_{0a}=\dot{\phi}\;I_a \und \mathcal{F}_{ab}=(\phi^2 -1)\,[I_a ,I_b]\,,
\end{align}
where the dot denotes $\partial_u$. Notice that the color-magnetic components $\mathcal{F}_{ab}$ lie in $\mathfrak{h}$ and the color-electric components $\mathcal{F}_{0a}$ lie in $\mathfrak{m}$. In particular, for $\mathrm{dim}(M)=4$, this implies that there are no (anti-)self-dual gauge fields which are $G$ invariant.

The dynamics of the gauge field is obtained from the Yang--Mills action
\begin{align}\label{eq:action}
S=\frac{1}{4\alpha}\int_M K(\mathcal{F}\wedge\ast\mathcal{F})=\frac{1}{8\alpha}\int_{\mathbbm{R}}\int_{G/H}K(\mathcal{F}_{\munu},\mathcal{F}^{\munu})\mathrm{dVol}\,.
\end{align}
To obtain the reduced equation of motion for $\phi(u)$, we can either solve the Yang--Mills equation $\de^{\mathcal{A}}\ast\mathcal{F}=0$ with $\Acal$ and $\Fcal$ given respectively by \eqref{eq:A-reduced} and \eqref{ansatz} or substitute the gauge field directly into the action \eqref{eq:action} and then extremize it, obtaining the reduced action
\begin{align}\label{reduced action}
S[\phi]=\mathrm{Vol}(G/H)\int_{\mathbbm{R}}\underbrace{\frac{1}{8\alpha}K(\mathcal{F}_{\munu},\mathcal{F}^{\munu})}_{\mathcal{L}(\phi,\dot{\phi})}\de u\,.
\end{align}
Since $G$\footnote{Or its component connected to the identity.} is semi-simple, connected and analytic in all cases considered, the equivalence of these two routes is guaranteed by the principle of symmetric criticality \cite{Palais:1979rca}. Hence, we will solve the gauge dynamics from the extremization of the reduced action. Before proceeding to the study of concrete cases, we remark that, from \eqref{ansatz} and from the orthogonality of the basis of generators, the structure of the reduced Lagrangian will be that of a one-dimensional Newtonian particle subject to a quartic potential, that is, $\mathcal{L}\sim \dot{\phi}^2 - V(\phi)$, with $V(\phi)$ being an even polynomial of degree four.

\section{\bf Symmetric Yang--Mills fields in foliated spacetimes}

In the context of cosmology, the homogeneous and isotropic behavior of the universe in large scales tends to favor an inflation scenario described by a scalar field. However, the fundamental fields in high-energy physics are mostly gauge and fermionic fields, with the only scalar in the standard model of particle physics being the Higgs field, whose potential is incompatible with the slow-roll required for inflation. It is in this context that inflationary models driven by a gauge field (gauge-flation) arise \cite{Maleknejad2012:1212.2921v3}. Indeed, it was first shown by Hosotani \cite{HOSOTANI198444} that non-Abelian gauge fields can be consistent with an homogeneous and isotropic space and suitable for cosmological purposes. Classical solutions to the Einstein--Yang--Mills coupled system in applications to cosmology were further pursued, not only for inflationary models but also for considerations on particle physics in the early universe (see, \textit{e.g.}, \cite{Galtsov:1991un} and \cite{friedan2020origin}). 

With future potential applications to cosmology in mind, in this Section, we will apply the CSDR scheme described in Sec.~\ref{sec2:YM-CSDR} to derive the dynamics of such Yang--Mills fields in maximally symmetric spacetimes of either open, flat, or closed natures.

\subsection{Application of CSDR to \texorpdfstring{$H^n$}{Hn}, \texorpdfstring{dS$_n$}{dSn}, and \texorpdfstring{AdS$_n$}{AdSn}}\label{sec:ApplyingCSDR}

We start our discussion with the rather general application of the CSDR scheme to the non-compact cosets
hyperbolic space $H^n$, de Sitter space dS$_n$ and anti-de Sitter space AdS$_n$. 
While cylinders over the latter two allow for more exotic spacetimes, the hyperbolic, as well as the spherical case
to be included shortly, yield the familiar cosmological spacetime topologies of open and closed universes. 
All three cosets can be realized as symmetric spaces via
\begin{align}
H^n &\cong \mathrm{SO}(1,n)/\mathrm{SO}(n)\,,\\
\text{dS}_n &\cong \mathrm{SO}(1,n)/\mathrm{SO}(1,n-1)\,, \ \text{and}\\
\text{AdS}_n &\cong \mathrm{SO}(2,n-1)/\mathrm{SO}(1,n-1)\,,
\end{align}
which are quotients of orthogonal groups of indefinite signatures. They all have irreducible representations $\mathrm{ad}(\mathfrak{h})|_{\mathfrak{m}}$, reducing the dynamics of the Yang--Mills system in the CSDR scheme to a single degree of freedom $\phi(u)$. Hence, we will treat the derivation of the reduced Lagrangians \eqref{reduced action} for the three cases in parallel for now.

From the definitions of the Killing form \eqref{killing}, the metric on the cylinder \eqref{metric}, and the reduced field strength \eqref{ansatz}, we obtain for the Lagrangian \eqref{reduced action}
\begin{equation}
\begin{aligned}
\mathcal{L}(\phi,\dot{\phi})&=\frac{1}{8\alpha}K(\mathcal{F}_{\munu},\mathcal{F}^{\munu})\\
&=\frac{1}{8\alpha}\left(2K(\mathcal{F}_{0a},\mathcal{F}_{0b})g^{00}g^{ab} + K(\mathcal{F}_{ma},\mathcal{F}_{nb})g^{mn}g^{ab}\right)\\
&=\frac{1}{8\alpha}\mathcal{D}_{n}\varepsilon_K \,\left( 2\tilde{\eta}(\mathcal{F}_{0a},\mathcal{F}_{0b})\tilde{g}^{00}\tilde{\eta}^{ab} + \tilde{\eta}(\mathcal{F}_{ma},\mathcal{F}_{nb})\tilde{\eta}^{mn}\tilde{\eta}^{ab}\right)\\
&=\frac{1}{8\alpha}\mathcal{D}_{n}\varepsilon_K \,\left(2\dot{\phi}^2 n\tilde{g}^{00}+(\phi^2-1)^2 \tilde{\eta}([I_a,I_b],[I_a ,I_b])\tilde{\eta}^{aa}\tilde{\eta}^{bb} \right)\\
&=\frac{1}{2\alpha}\mathcal{D}_{n}n\varepsilon_K \tilde{g}^{00}\,\left(\frac{1}{2}\dot{\phi}^2 + \frac{\tilde{g}^{00}}{4n}\mathcal{S}_{n}(\phi^2- 1)^2 \right)\,,
\end{aligned}
\end{equation}
where we have used $\tilde{\eta}^{ab}\tilde{\eta}_{ab}=\mathrm{dim}(\mathfrak{m})=n$. Thus, as previously anticipated, the Lagrangians reduce to that of a Newtonian particle subject to a quartic potential whose coefficient depends on the specific spacetime,
\begin{align}
V(\phi)=-\frac{\tilde{g}^{00}}{4n}\mathcal{S}_{n}\,(\phi^2 -1)^2\,,
\end{align}
where we defined
\begin{align}\label{double sum}
\mathcal{S}_{n}:=\tilde{\eta}([I_a,I_b],[I_a ,I_b])\tilde{\eta}^{aa}\tilde{\eta}^{bb}=\sum_{I,J\in\mathfrak{m}}\lVert [I,J] \rVert^{2}_{\tilde{\eta}}\,\lVert I \rVert^{2}_{\tilde{\eta}}\,\lVert J \rVert^{2}_{\tilde{\eta}}\,.
\end{align}
To obtain the explicit reduced Lagrangians for the three cases, we simply have to calculate the Killing forms, which fix $\tilde{g}_{00}$, then evaluate the double sum \eqref{double sum}. We begin with the Killing forms. Working in the defining matrix representations, we use 
\begin{align}
\mathrm{tr}_{\mathrm{adj}}(XY) =(p+q-2) \mathrm{tr}_{\mathrm{def}}(XY)\,,
\end{align}
which holds on $SO(p,q)$ with $p+q\geq 3$, $p,q\geq 1$. Then, we get
\begin{align}
H^n \;:\;\;\; &\tilde{\eta}_{AB} = \tilde{\eta}_{\alpha\beta}\oplus \tilde{\eta}_{ab}= 
-\mathbb{1}_{\binom{n}{2}}
\oplus \mathbb{1}_{n}\,,\\
\text{dS}_n \;:\;\;\; &\tilde{\eta}_{AB} = \tilde{\eta}_{\alpha\beta}\oplus \tilde{\eta}_{ab}=\begin{pmatrix}
\mathbb{1}_{n-1} & \\
& - \mathbb{1}_{\binom{n-1}{2}}
\end{pmatrix}\oplus\begin{pmatrix}
1 & \\
& - \mathbb{1}_{n-1}
\end{pmatrix}\,, \quad \text{and}\\ \label{ads}
\text{AdS}_n \;:\;\;\; &\tilde{\eta}_{AB} = \tilde{\eta}_{\alpha\beta}\oplus \tilde{\eta}_{ab}=\begin{pmatrix}
\mathbb{1}_{n-1} & \\
& - \mathbb{1}_{\binom{n-1}{2}}
\end{pmatrix}\oplus\begin{pmatrix}
\mathbb{1}_{n-1} & \\
& -1
\end{pmatrix}\,,
\end{align}
with the Killing-normalization \eqref{killing} being $\mathcal{D}_{n} =2(n-1)$ in all scenarios. Naturally, the de Sitter case is just a reordering of the hyperbolic case. Now, from the coset part of these expressions, we see that $\tilde{g}_{00}$ is $-1$ for both $H^n$ and dS$_n$ and $+1$ for AdS$_n$, making the preliminary metric components $\tilde{g}_{\munu}$ on the cylinders \eqref{metric} mostly plus, mostly minus and mostly plus, respectively, which can be changed from $\tilde{g}$ to $g$ using the overall factor $\varepsilon_g$. We now evaluate the double sum \eqref{double sum}. We avoid working with structure constants explicitly and resort to combinatoric arguments. The main issue encountered in the sum is the indefiniteness of the Killing forms. Since commutators are involved, one has to keep track of the three signs arising when summing over all generators, thus, we simply look at all possible combinations that can occur. To this end, let us call generators which have $\tilde{\eta}$-square $-1$ `compact' $(C)$ and those with $\tilde{\eta}$-square $+1$ `non-compact' $(\neg C)$\footnote{Naturally, this definition lines up with the generators being (anti-)symmetric (as matrices) in the defining representation .}. Generally, we have
\begin{equation}\label{commutator1}
\begin{aligned}
[C,C]&=C\;\;\;\;\;\;\;\;\;\;\;\; [-,-]=-\\
[C,\neg C]&=\neg C\;\;\Leftrightarrow\;\; [-,+]=+\\
[\neg C,\neg C]&=C \;\;\;\;\;\;\;\;\;\;\;\; [+,+]=-
\end{aligned}\;.
\end{equation}
Hence, for the Lorentz algebra $\mathfrak{so}(1,n)$ and $I \neq J$, we have
\begin{table}[hbt!]
\centering
 \begin{tabular}{||c| c| c| c||} 
 \hline
 $\,\lVert [I,J] \rVert_{\tilde{\eta}}^{2}\,$ & $\,\lVert I \rVert_{\tilde{\eta}}^{2}\,$ & $\,\lVert J \rVert_{\tilde{\eta}}^{2}\,$ & $\,\Pi\,$ \\ [0.5ex] 
 \hline\hline
 $-$ & $-$ & $-$ & $-$ \\ \hline
 $+$ & $+$ & $-$ & $-$ \\ 
 $+$ & $-$ & $+$ & $-$ \\ \hline
 $-$ & $+$ & $+$ & $-$ \\ [0.4ex] 
 \hline
 \end{tabular}\;,
\end{table}\\
where $\Pi = \lVert [I,J] \rVert_{\tilde{\eta}}^{2} \lVert I \rVert_{\tilde{\eta}}^{2} \lVert J \rVert_{\tilde{\eta}}^{2}$ is the summand, which is $-1$ for all non-trivial ($I \neq J$) cases. With this, we obtain $\mathcal{S}_n =- 2\binom{n}{2}$ for the hyperbolic space and for the de Sitter space, which in turn yields
\begin{align}
V_{H^n}(\phi)=V_{\text{dS}_n}(\phi)=-\sf{1}{8}\mathcal{D}_{n}(\phi^2-1)^2\,,
\end{align}
where we have used $2\binom{n}{2}/(4n)=(n-1)/4=\mathcal{D}_{n}/8$. We proceed analogously for the anti-de Sitter space. We obtain the non-trivial combinations of $\mathfrak{so}(2,n-1)$ as 
\begin{table}[hbt!]
\centering
 \begin{tabular}{||c| c| c| c||} 
 \hline
 $\lVert [I,J] \rVert_{\tilde{\eta}}^{2}$ & $\lVert I \rVert_{\tilde{\eta}}^{2}$ & $\lVert J \rVert_{\tilde{\eta}}^{2}$ & $\Pi$ \\ [0.5ex] 
 \hline\hline
 $+$ & $+$ & $-$ & $-$ \\ 
 $+$ & $-$ & $+$ & $-$ \\ \hline
 $-$ & $+$ & $+$ & $-$ \\ [0.5ex] 
 \hline
 \end{tabular}\,.
\end{table}\\
There is no non-trivial case with both $I$ and $J$ $\tilde{\eta}$-squaring to $-1$ since here there is only one compact generator, which is clear from \eqref{ads}. The sum becomes trivial, as in the previous cases, but this time with $\tilde{g}_{00}=+1$, yielding
\begin{align}
V_{\text{AdS}_n}(\phi)=+\sf{1}{8}\mathcal{D}_{n}(\phi^2-1)^2\,.
\end{align}

With the potentials computed explicitly, we have obtained the reduced Lagrangians for the three cases at hand. The potentials for the quasi-Newtonian degrees of freedom $\phi(u)$ are the inverted double well for the hyperbolic and the de Sitter space and the usual double well for the anti-de Sitter space. For $n=3$, it matches previous results in \cite{Mink}, where the special cases $H^3$ and dS$_3$ were discussed. Not only do the three systems behave similarly but the Lagrangians and potentials are all structurally identical, with
\begin{align}\label{lagrangian}
\mathcal{L}=\frac{1}{2\alpha}\varepsilon_{K}\tilde{g}_{00}\,n\,\mathcal{D}_{n}\left(\frac{1}{2}\dot{\phi}^2 -V(\phi)\right) \und V(\phi)=\tilde{g}_{00}\frac{1}{8}\mathcal{D}_n (\phi^2 -1)^2\,,
\end{align}
where the only difference is the presence of $\tilde{g}_{00}$ in the more comprehensive treatment shown here. Furthermore, the same discussion applies analogously to the case SO$(n+1)$/SO$(n)\cong S^n$ discussed in \cite{Lechtenfeld_2018}, which completes the picture with the positive curvature-dual of the hyperbolic space. Translating the spheric case into our notation amounts to setting $\varepsilon_K=+1$, $\varepsilon_g=-1$ and $\tilde{g}_{00}=+1$. Hence, we include it into our picture and arrive at the following relation:
\begin{table}[hbt!]
\centering
 \begin{tabular}{c c c} 
 $V_{S^n}$ & $=$ & $-V_{H^n}$ \\ 
 $\veq$ & & $\veq$ \\ 
 $V_{\text{AdS}_n}$ & $=$ & $-V_{\text{dS}_n}$ \\ 
 \end{tabular}
\end{table}

\subsection{Energy-momentum tensors}

Similarly as for the reduced Lagrangian, we can compute closed expressions for the energy-momentum tensors of all cases here considered at once, including the sphere, as mentioned above. We use the standard, divergence-free and gauge invariant Yang--Mills energy-momentum tensor given by
\begin{align}
T_{\munu}=-\frac{1}{2\alpha} \left( K\left(\mathcal{F}_{\mu\sigma},\mathcal{F}_{\nu\rho}\right)g^{\sigma\rho} - \frac{1}{4}g_{\munu}K(\mathcal{F}_{\ab},\mathcal{F}^{\ab})\right)\;,\;\; T=T_{\munu}\,e^\mu \otimes e^\nu \,.
\end{align}
The calculation is straightforward and relies on similar combinatoric considerations as for the potentials. Let us begin with the first term of $T_{\munu}$. The 00-component is
\begin{align}
K(\mathcal{F}_{0\sigma},\mathcal{F}_{0\rho})g^{\sigma\rho}=n\mathcal{D}_n \varepsilon_K \varepsilon_g \dot{\phi}^2\,. 
\end{align}
For the coset components, we get
\begin{equation}\label{cosetcomponents}
\begin{aligned}
K(\mathcal{F}_{a\sigma},\mathcal{F}_{b\rho})g^{\sigma\rho} &= \varepsilon_g \tilde{g}^{00} K(\mathcal{F}_{a0},\mathcal{F}_{b0})+K(\mathcal{F}_{am},\mathcal{F}_{bn})g^{mn}\\
&=\varepsilon_g \tilde{g}^{00}\varepsilon_K \mathcal{D}_n \tilde{\eta}_{ab} \dot{\phi}^2 + (\phi^2 -1)^2 \mathcal{D}_n \varepsilon_K \tilde{\eta}\left([I_a ,I_m ],[I_b , I_n ]\right) \varepsilon_g \tilde{\eta}^{mn}.
\end{aligned}
\end{equation}
We again encounter a term which sums over products of $\tilde{\eta}$-squares,
\begin{equation}
\begin{aligned}
&\tilde{\eta}\left([I_a ,I_m ],[I_b , I_n ]\right) \tilde{\eta}^{mn}=\sum_{I\in\mathfrak{m}}\tilde{\eta}\left( [I, I_a ] , [I, I_b ] \right)\,\lVert I \rVert_{\tilde{\eta}}^{2} \\
&=\sum_{I\in\mathfrak{m}}\lVert [I, I_a ] \rVert_{\tilde{\eta}}^{2}\,\lVert I \rVert_{\tilde{\eta}}^{2}\,\delta_{ab} =:\mathcal{C}_a \,\delta_{ab}\,,
\end{aligned}
\end{equation}
where in the last line we have used that $\mathrm{ad}(I)\equiv [I,\cdot]:\{I_a\}\rightarrow\{I_\alpha\}$ is injective. Again, we perform a sign analysis. From our previous computations, we know that the relations \eqref{commutator1} hold for all cases, spheres included (which only have compact generators). We thus consider all combinations:
\begin{table}[hbt!]
\centering
 \begin{tabular}{||c| c| c| c||} 
 \hline
 $\lVert [I_a] \rVert_{\tilde{\eta}}^{2}$ & $\lVert I \rVert_{\tilde{\eta}}^{2}$ & $\lVert [I,I_a] \rVert_{\tilde{\eta}}^{2}$ & $\lVert I \rVert_{\tilde{\eta}}^{2}\,\lVert [I,I_a] \rVert_{\tilde{\eta}}^{2}$ \\ [0.5ex] 
 \hline\hline
 $+$ & $+$ & $-$ & $-$ \\ 
 $+$ & $-$ & $+$ & $-$ \\ \hline
 $-$ & $+$ & $+$ & $+$ \\ 
 $-$ & $-$ & $-$ & $+$ \\ [0.5ex] 
 \hline
 \end{tabular}\,.
\end{table}\\ 
For a fixed $a$, the summand is either always $-1$ or $+1$, except for the trivial $I=I_a$ case where it is zero. So, we get
\begin{align}
\mathcal{C}_{a}=\begin{cases}
-(n-1)\,,\; a\text{ is }\neg C\\
+(n-1)\,,\; a\text{ is } C
\end{cases} ,
\end{align}
and, combining $\mathcal{C}_a$ with $\delta_{ab}$, we obtain
\begin{align}
\mathcal{C}_{a}\delta_{ab}=-(n-1)\tilde{\eta}_{ab}\,.
\end{align}
With that, equation \eqref{cosetcomponents} can be further simplified, yielding
\begin{equation}
\begin{aligned}
K(\mathcal{F}_{a\sigma},\mathcal{F}_{b\rho})g^{\sigma\rho} &=\varepsilon_g \tilde{g}^{00}\varepsilon_K \mathcal{D}_n \tilde{\eta}_{ab} \dot{\phi}^2 - (\phi^2 -1)^2 \mathcal{D}_n \varepsilon_K \varepsilon_g (n-1)\tilde{\eta}_{ab}\\
&=\varepsilon_K \varepsilon_g \mathcal{D}_n \tilde{\eta}_{ab} \tilde{g}_{00} \left(\dot{\phi}^2 -4 V(\phi)\right)\,,
\end{aligned}
\end{equation}
where in the last step we have used $\tilde{g}_{00}^2 =1$. The other mixed components of the first term of $T_{\munu}$ all vanish, since
\begin{align}
 K\left(\mathfrak{h},\mathfrak{m}\right)\equiv 0\,.
\end{align}
The second term in the energy-momentum tensor is proportional to the Lagrangian itself. Hence, combining both terms, we obtain the energy-momentum tensor explicitly:
\begin{equation}\label{Tmunu}
\begin{aligned}
T = -\frac{\varepsilon_K \varepsilon_g \mathcal{D}_n}{2\alpha} \left( n\left(\frac{1}{2}\dot{\phi}^2 + V\right)e^0 \otimes e^0 + \tilde{g}_{00} \left( \left(1-\frac{n}{2}\right)\dot{\phi}^2 +(n-4) V \right)\tilde{\eta}_{ab}\, e^a \otimes e^b \right).
\end{aligned}
\end{equation}

Notice that the sign of $T$ depends both on the overall sign of the metric $\varepsilon_g$, which is expected, and also on the overall sign of the Killing form $\varepsilon_K$.

For the particular cases of Riemannian cosets $S^n$ and $H^n$, the spacetimes are homogeneous and isotropic. In this case, the spatially homogeneous Yang--Mills fields yield perfect fluid energy-momentum tensors. This is not true anymore for the slicings with Lorentzian cosets, (A)dS$_n$. Moreover, since the foliation parameter is spacelike in the latter, the energy density will not reside in $T_{00}$, but in $T_{aa}$ for the timelike direction $a$. For $n=3$, that is, four dimensional spacetime, the energy-momentum tensors read
\begin{align}\label{fluid}
T=-\varepsilon_K \varepsilon_g \frac{6 E}{\alpha} \,\left( e^0 \otimes e^0 - \frac{1}{3} \tilde{g}_{00}\, \tilde{\eta}_{ab}\, e^a \otimes e^b \right), 
\end{align}
where we have defined the mechanical ``energy" of the analog particle
\begin{align}\label{MechEn}
E := \frac{1}{2}\dot{\phi}^2 + V(\phi).
\end{align}
The results again match \cite{Kumar2022:2206.12009v1} for hyperbolic and de Sitter spaces and \cite{Kumar_2021} for the sphere. The energy-momentum tensors are traceless and, for the Riemannian slicings\footnote{where $\tilde{g}_{00}\,\tilde{\eta}_{ab}=-\delta_{ab}$ holds} with $S^3$ and $H^3$, they are of the perfect-fluid radiation type, as expected.

\subsection{Solutions to equation of motion}
The reduced equations of motion are that of an one-dimensional Newtonian particle subject to either a double-well or inverted double-well potential. Alternatively, we can solve the Newtonian dynamics of $\phi$ from the conservation of energy, with
\begin{align}
\sf{1}{2}\dot{\phi}^2 \pm \sf{1}{8}\mathcal{D}_{n}(\phi^2 -1)^2 =E\,.
\end{align}
The potential scales with the dimension $n$ of the coset, as $\mathcal{D}_{n}=2(n-1)$. We now solve the dynamics analytically for both the double well and the inverted double well cases.

\begin{figure}[hbt!]
\includegraphics[scale=0.55]{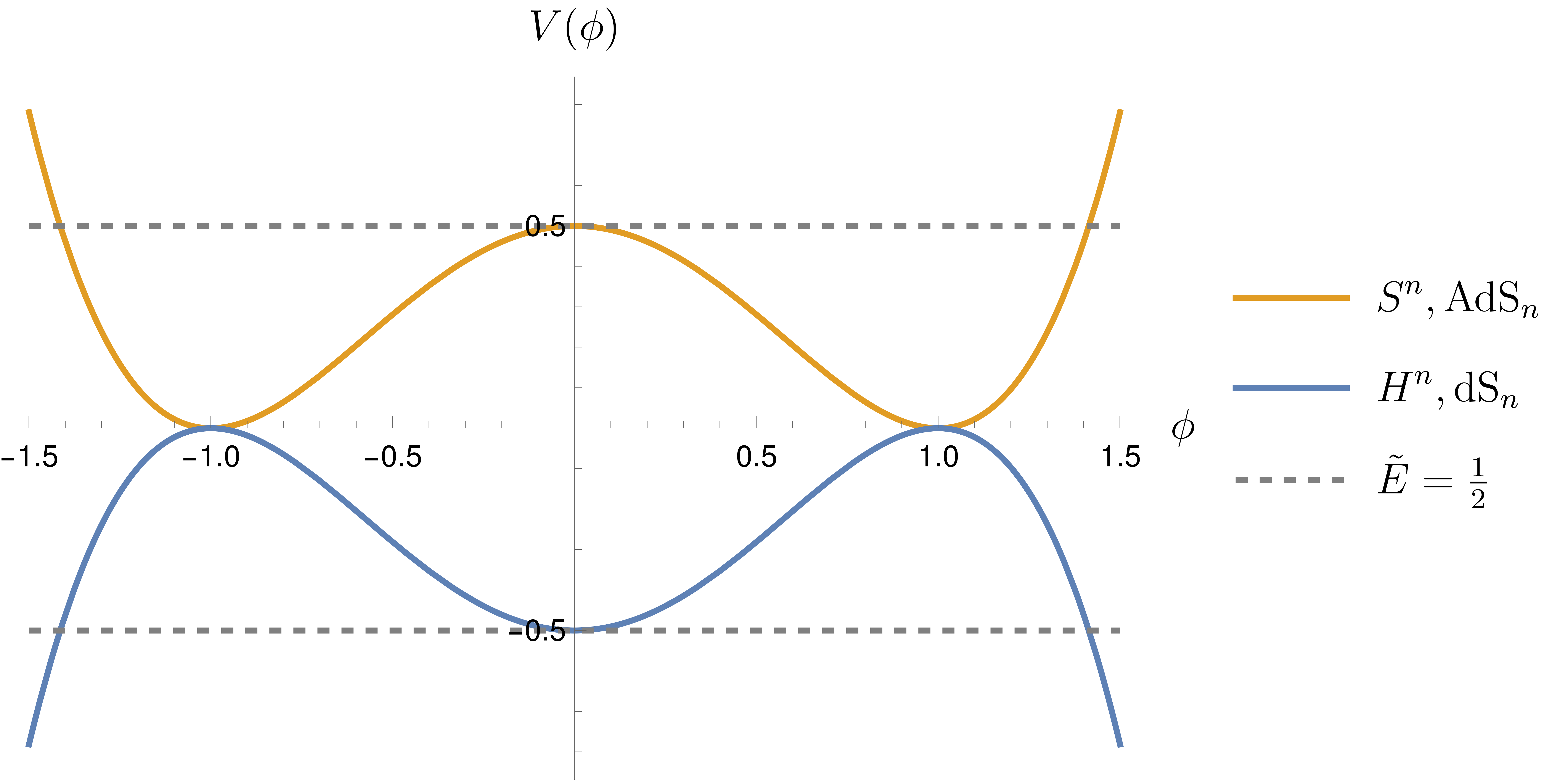}
\caption{Plots of mechanical analogue potentials for $n=3$ constructed with the CSDR scheme on symmetric spaces over orthogonal groups.}
\label{figPots}
\end{figure}

Using time translation invariance to fix $\dot{\phi}(u=0)=0$, the solutions are parametrized by the overall energy $E$ and by the initial value $\phi_{\ast}:=\phi(u=0)$. For the double well $V(\phi)=+\frac{1}{8}\mathcal{D}_{n}(\phi^2-1)^2$ and $E\geq 0$, the solutions are
\begin{align}\label{eq:Jacobi-phi-straight}
\phi(u)=\pm\begin{cases}
\frac{k}{\varepsilon}\mathrm{cn}\left(\frac{\sqrt{\mathcal{D}_{n}}}{2\varepsilon}\,u, k^2 \right)\;&,\;\tilde{E}\in(0,\infty) \leftrightarrow k^2 \in (\infty,\frac{1}{2})\\
0\;&,\;\; \tilde{E}=\frac{1}{2}\leftrightarrow k^2 =1 \\
1 \;&,\;\; \tilde{E}=0\leftrightarrow k^2 =\infty
\end{cases}\ \,,
\end{align}
where 
\begin{align}
\tilde{E}=4\frac{E}{\mathcal{D}_{n}}\geq 0\;,\;\; \varepsilon^2 = \frac{1}{2\sqrt{2\tilde{E}}}\;,\;\; k^2 =\frac{1}{2}+\varepsilon^2\,,
\end{align}
and $\mathrm{cn}$ denotes the Jacobi elliptic function. For values of $\tilde{E}\in (\frac{1}{2},\infty)$, the field oscillates through the whole potential with average value $0$, while, for $\tilde{E}\in (0,\frac{1}{2})$, it acquires a non-trivial average value close to one of the two minima of the double well potential $V(\phi)$ and becomes restricted to the corresponding well, see Fig.~\ref{figPots}. For $\tilde{E}=\frac{1}{2} \leftrightarrow k^2=1$, the first solution in \eqref{eq:Jacobi-phi-straight} simplifies to 
\begin{equation}
	\phi(u) = \pm \sqrt{2}\operatorname{cn}\left( \sqrt{\frac{\mathcal{D}_n}{2}} u , 1 \right) = \pm \sqrt{2}\operatorname{sech}\left( \sqrt{\frac{\mathcal{D}_n}{2}} u \right)\,,
\end{equation}
which is the so-called bouncing solution, with $\phi(u)\rightarrow 0$ for $u\rightarrow \pm\infty$. Additionally, we also have three trivial solutions on the local extrema of the potential. When $\tilde{E}=\frac{1}{2} \leftrightarrow k^2=1$, there is the unstable trivial solution $\phi(u)=0$, corresponding to the Maurer-Cartan form of $H$, $\mathcal{A}=I_\alpha e^\alpha$, rendering the field strength purely magnetic. When $\tilde{E}=0 \leftrightarrow k^2=\infty$, there are the two stable solutions $\phi(u) \equiv \pm 1$, which correspond to the two gauge-equivalent versions of the Maurer-Cartan form of $G$, $\mathcal{A}= I_\alpha e^\alpha \pm I_{a} e^a$, and render the connection flat, \textit{i.e.} $\mathcal{F}\equiv 0$.

Solutions for the inverted potential $V(\phi)=-\frac{1}{8}\mathcal{D}_{n}(\phi^2-1)^2$ and $E\leq 0$ are obtained via a Wick rotation of the non-inverted case\footnote{Or equivalently $\mathcal{D}_n \mapsto -\mathcal{D}_n$.}:
\begin{align}\label{eq:Jacobi-phi-inverted}
\phi(u)=\pm \begin{cases}
\frac{k}{\varepsilon}\operatorname{nc}\left(\frac{\sqrt{\mathcal{D}_{n}}}{2\varepsilon}\,u,1- k^2 \right)\;&,\;\;\tilde{E}\in(0,\infty) \leftrightarrow k^2 \in (\infty,\frac{1}{2})\\
\frac{\sqrt{k^2-1}}{\varepsilon}\operatorname{cd}\left(\frac{k\sqrt{\mathcal{D}_n}}{2\varepsilon}\,u,1- \frac{1}{k^2} \right)\;&,\;\;\tilde{E}\in(0,\frac{1}{2}) \leftrightarrow k^2 \in (\infty,1)\\
\operatorname{tanh}\left( \frac{\sqrt{\mathcal{D}_n}}{2}\,u \right)\;&,\;\; \tilde{E}=0\leftrightarrow k^2 =\infty\\
0\;&,\;\; \tilde{E}=\frac{1}{2}\leftrightarrow k^2 =1 \\
1 \;&,\;\; \tilde{E}=0\leftrightarrow k^2 =\infty
\end{cases}\ \,,
\end{align}
where
\begin{align}
\tilde{E}=-4\frac{E}{\mathcal{D}_n}\geq 0 \;,\;\; \varepsilon^2 =\frac{1}{2\sqrt{2\tilde{E}}}\;,\;\; k^2 =\frac{1}{2}+\varepsilon^2 \,,
\end{align}
and nc and cd again denote Jacobi elliptic functions. The first solution in \eqref{eq:Jacobi-phi-inverted} corresponds to runaway solutions that diverge at finite time $u$. The second solution corresponds to periodic functions inside the well, with $\tilde{E}\in(0,\frac{1}{2})$. When $\tilde{E}\rightarrow 0$, there is an extra solution in which we cannot impose the $\dot{\phi}(0)=0$ condition, namely the third function in \eqref{eq:Jacobi-phi-inverted}, which is usually called the \textit{meron} or \textit{sphaleron}. Lastly, as in the straight potential, there are the same three trivial solutions at the local extrema corresponding to the Maurer-Cartan forms, at $\tilde{E}=\frac{1}{2}$ or $\tilde{E}=0$.

\section{\bf Spacetime dynamics: the Einstein--Yang--Mills coupled system}\label{sec3:GR}

In this section, we will consider dynamical spacetimes in Friedmann--Lemaître--Robertson--Walker universes of open, flat, or closed nature. We will investigate the coupled Einstein--Yang--Mills system to probe the effects of $G$-invariant gauge fields on the cosmological evolution of the universe, and vice-versa.

\subsection{FLRW dynamics in \texorpdfstring{$n{+}1$}{n+1} dimensions}\label{sec3:GRA}

The dynamics of homogeneous and isotropic spacetimes are reduced to that of a single scalar $a(t)$, the scale factor, with the general FLRW metric reading
\begin{align}
g=-\de t\otimes\de t + a^2(t) g_{\tilde{M}}\,.
\end{align}
The spatial metric $g_{\tilde{M}}$ is the `canonical', SO$(n)$-invariant Riemannian metric of constant (normalized) sectional curvature for $\tilde{M}=S^n,H^n,\mathbbm{R}^n$ corresponding respectively to $\mathrm{sec}\equiv k =\pm 1 ,0$. The spacetime dynamics is, in general, dictated by the Einstein equations,
\begin{align}
\mathcal{G}+\Lambda g &= \kappa T \;, \quad \text{where}\quad \mathcal{G}=Ric -\sf{1}{2}\mathcal{R}\,g
\end{align}
and $\kappa$ is the gravitational coupling. Homogeneous and isotropic matter has its energy-momentum tensor reduced to the perfect-fluid type,
\begin{equation}
 	T^{\mu\nu} = (\rho+p)u^\mu u^\nu + p g^{\mu\nu}\,,
\end{equation} 
for some $(n{+}1)$-velocity $u$ for the fluid. In this case, the Einstein equations are reduced to the Friedmann equations for $a(t)$:
\begin{align}
	\frac{n(n{-}1)}{2}\left( \frac{k}{a^2} + \left(\frac{\dot{a}}{a}\right)^2 \right) - \Lambda &= \kappa\,\rho\,,\\
	-(n{-}1)\frac{\ddot{a}}{a} - \frac{(n{-}2)(n{-}1)}{2}\left( \frac{k}{a^2} + \left(\frac{\dot{a}}{a}\right)^2 \right) + \Lambda &= \kappa\,p\,.
\end{align}
Ultimately, the evolution of the scale factor depends on the matter content considered and its equation of state. In Sec.~\ref{sec2:YM-CSDR}, we showed that the energy-momentum tensors of $G$-invariant gauge fields obtained from the CSDR scheme are of the perfect-fluid type and we obtained $\rho$ and $p$, according to equation \eqref{Tmunu}.
Due to the conformal invariance of gauge theories in four-dimensional spacetimes, the full system reduces to a one-way coupling, \textit{i.e.} the Yang--Mills equations are independent of the scale factor. However, in the general $(n+1)$-dimensional case scenario, both the equation of state and the equation of motion of the gauge field become more involved. The general case will be considered in Sec.~\ref{sec4:warping}. For now, let us turn our attention to the exactly solvable case of four dimensional spacetimes that are foliated with the Riemannian slices, namely $S^3$ and $H^3$.

It is often more useful to parametrize the time evolution in terms of the conformal time $\tau(t)$, with
\begin{align}
\de \tau = \frac{1}{a}\de t \;\Rightarrow\; \tau (t)=\int \de t \frac{1}{a(t)}\,,
\end{align}
such that the metric becomes explicitly conformally flat:
\begin{align}
g=a^2 (\tau )\left(-\de\tau\otimes\de\tau +g_{\tilde{M}}\right)\,.
\end{align}
Identifying the foliation parameter $u$ in the Yang--Mills construction with the conformal time, the energy-momentum tensors \eqref{fluid} become
\begin{align}
T=\varepsilon_K \tilde{g}_{00}\frac{6 E}{a(\tau)^2 \,\alpha}\left(\de\tau\otimes\de\tau + \frac{1}{3} g_{\tilde{M}}\right)\,,
\end{align}
where we have used $\varepsilon_g = -\tilde{g}_{00}$, which holds for the Riemannian cosets while working explicitly in mostly plus signature. Note that, in the expression above, the overdots in the definition of $E$, see \eqref{MechEn}, are now to be read as derivatives with respect to the conformal time. In this case, as previously mentioned, the coupled Einstein--Yang--Mills system reduces to a one-way coupling and is analytically solvable. This idea has been explored, for example, in \cite{friedan2020origin,Kumar_2021} for the FLRW-type closed universe, \textit{i.e.} the $S^3 \cong$ SU$(2)$ case. We now discuss the general equations in four dimensions, which include the aforementioned closed-universe solutions, and then particularize to the open universe, with $k=-1$.

\subsection{Analytic solutions in 4D spacetimes with \texorpdfstring{$G$}{G}-invariant Yang--Mills fields}

\begin{figure}[t]
\includegraphics[scale=0.48]{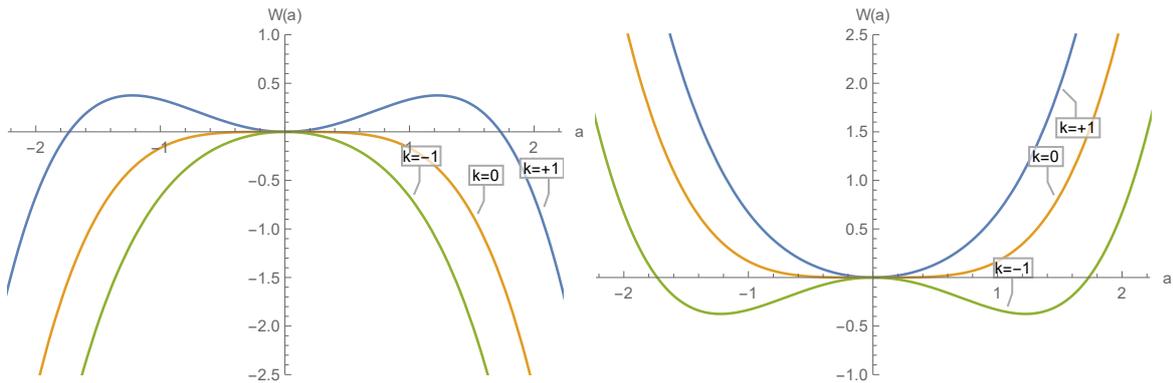}
\caption{Plots of the cosmological potential $W(a)$ for different spacetime topologies $k=\pm 1, 0$ and cosmological constants $\Lambda = +1$ (left), $\Lambda = -1$ (right).}
\label{figW}
\end{figure}

Using $\tr_g(T)=0$ in four dimensions further simplifies the Friedmann equations, which are then reduced to
\begin{align}\label{eq:reduced-grav-eqns}
\ddot{a}+W'(a)&=0 \\
E_{\GR} := \sf{1}{2}\dot{a}^2+W(a)&=\sf{1}{6}\kappa T_{00}\,,
\end{align}
with the dot denoting differentiation with respect to the conformal time $\tau$. The cosmological potential $W(a)$, illustrated in Fig.~\ref{figW}, and hence the range of possible dynamics of the spacetime, is determined by the signs of the sectional curvature $k$ of the spatial slicing and of the cosmological constant $\Lambda$,
\begin{align}
W(a)=\frac{k}{2}a^2 -\frac{\Lambda}{6}a^4\,.
\end{align}
The equations of motion for the scale factor in this case are, like for the Yang--Mills system, that of a Newtonian particle subject to a quartic cosmological potential. The coupling to the Yang--Mills system is given solely through the energy balance
\begin{equation}\label{energybalancing}
\begin{aligned}
\frac{1}{2}\dot{a}^2 +W(a) &= \frac{\kappa}{\alpha} \varepsilon_K \tilde{g}_{00} \left(\frac{1}{2}\dot{\phi}^2 +V(\phi)\right)\\
\Leftrightarrow E_{\mathrm{GR}}&=\frac{\kappa}{\alpha}\varepsilon_K \tilde{g}_{00}\, E_{\mathrm{YM}}\,,
\end{aligned}
\end{equation}
known as the Wheeler--DeWitt constraint. In \cite{friedan2020origin,Kumar_2021}, where the case $k=+1 \leftrightarrow \tilde{M}=S^3 \cong \text{SU}(2)$ with $\Lambda >0$ was considered, the Yang--Mills system was subject to a double well and the scale factor to an inverted double well, resulting either in periodic or in blow-up solutions for $a(\tau )$. 

We now consider the case $k=-1 \leftrightarrow \tilde{M}=H^3$, \textit{i.e.} open-type hyberbolic cosmologies. In this case, the Yang--Mills system is subject to an inverted double well and the dynamics of the scale factor depends on the sign of $\Lambda$: 
\begin{itemize}
\item[(i)] $\Lambda>0\;\Rightarrow\; W(a)$ is concave $\;:\;$ $E_{\mathrm{GR}}\neq 0$ yields only blow-up solutions and $E_{\mathrm{GR}}=0$ yields one trivial solution $a\equiv 0 $ and `big crunch' scenarios.
\item[(ii)] $\Lambda<0\;\Rightarrow\; W(a)$ double well $\;:\;$ Oscillatory solutions both with and without `big crunch' and stationary solutions at the potential's extrema.
\end{itemize}
Let us focus on the $\Lambda <0$ case. Fixing $\dot{a}(0)=0$, the solutions for the scale factor in conformal time can be parametrized by the initial energy $E_{\mathrm{GR}}$ and initial condition $a(0)$. Similarly to \eqref{eq:Jacobi-phi-straight}, we have
\begin{align}
a(\tau)=\pm\begin{cases}
\sqrt{\frac{-3}{2\Lambda}}\frac{k}{\varepsilon}\mathrm{cn}\left(\frac{1}{\sqrt{2}\varepsilon}\, \tau , k^2 \right) \;&,\;\; E_{\GR} \in (\frac{3}{8\Lambda},\infty)\leftrightarrow k^2 \in(\infty,\frac{1}{2}) \\
0 \;&,\;\; E_{\GR} = 0 \leftrightarrow k^2 =1 \\
\sqrt{\frac{-3\Lambda}{2}} \;&,\;\; E_{\GR} =\frac{3}{8\Lambda}\leftrightarrow k^2 =\infty \\
\end{cases}\ \,,
\end{align}
with 
\begin{align}
E_{\GR}\in\left(\frac{3}{8\Lambda},\infty\right)\;,\;\;\varepsilon^2 = \frac{1}{2\sqrt{1-\frac{8\Lambda}{3} E_{\GR}}}\,,\und k^2 = \frac{1}{2}+\varepsilon^2\,.
\end{align}
Once more, the first solution has three different qualitative behaviours. For $E_{\mathrm{GR}}>0$, the scale factor evolves from $0$ to some maximum value, then returns to $0$ at finite time. For $E_{\mathrm{GR}}<0$, the scale factor oscilattes with finite period between a maximum and a minimum value, never reaching $0$. For $E_{\mathrm{GR}}=0$, the scale factor is a maximum at $\tau=0$ and asymptotically goes to $0$ in both past and future infinities. Additionally, there are the three usual trivial solutions at the three local extrema.

\begin{figure}[hbt!]
\centering
\includegraphics[scale=0.48]{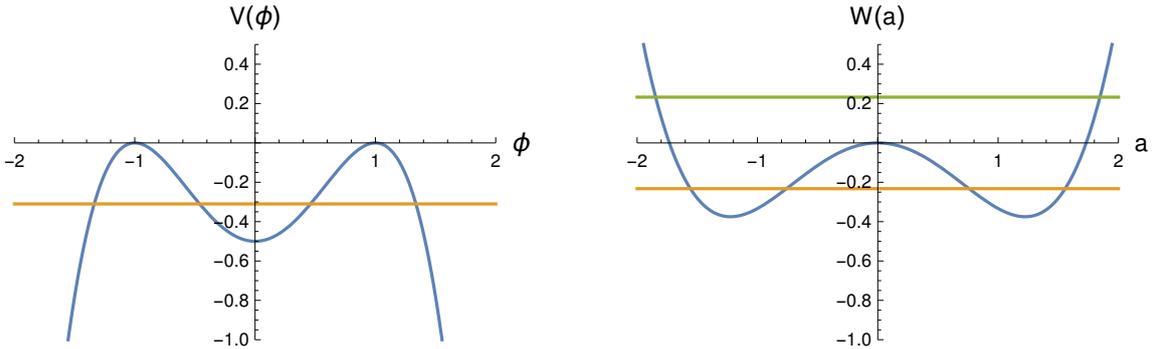}
\caption{Example of energy balancing of the Yang--Mills field and the scale factor depending on the choice of $\varepsilon_K = \pm\tilde{g}_{00}$ for $\Lambda = -1$, $\alpha=\frac{4}{3}$ (minimal) and $\kappa =1$.}
\label{figVandW}
\end{figure}

Going back to the energy-balance condition \eqref{energybalancing}, we notice that there is a freedom to choose the sign $\varepsilon_K$ with which the energy of the analog particle $\phi$, $E_{\mathrm{YM}}$, couples to the spacetime dynamics through $E_{\mathrm{GR}}$. Indeed, for the bounded solutions $\phi(\tau)$ of the inverted double well, the energy is always non-positive, with bounded solutions around the local minima for the scale factor. More precisely, we have
\begin{align}
E_{\mathrm{YM}}\in\left[-\sf{1}{2},0\right] \;\;\Leftrightarrow\;\; \frac{\alpha}{\kappa}\varepsilon_K \tilde{g}_{00}\,E_{\mathrm{GR}}\in \left[-\sf{1}{2},0\right]\,,
\end{align}
yielding two possibilities
\begin{center}
\begin{itemize}
\item[(i)] $\varepsilon_K =\tilde{g}_{00}$ $\rightarrow$ energy couples directly $E_{\mathrm{GR}}\in\left[-\frac{\kappa}{2\alpha},0\right]$\,, \\
\item[(ii)] $\varepsilon_K =-\tilde{g}_{00}$ $\rightarrow$ energy couples invertedly $E_{\mathrm{GR}}\in\left[0,\frac{\kappa}{2\alpha}\right]$,
\end{itemize}
\end{center}
both of which are illustrated in Fig.~\ref{figVandW}.
Notice that any bounded configuration for $\phi$, with $0>E_\mathrm{YM}>-\frac{1}{2}$, provides a physical region for the cosmological scale factor in scenario (ii), where big bag initial conditions are always possible and a \textit{big crunch} is inevitable. The same is not true for scenario (i), in which all physical solutions for the scale factor oscillate around a non-zero value and never reach zero. If we are to allow for the Yang-Mills field to sit at its local minimum $E_\mathrm{YM}=-\frac{1}{2}$, we get a constraint involving the gauge coupling, the gravitational coupling, and the cosmological constant such that there are physical solutions for the scale factor, namely:
\begin{align}
\frac{\alpha}{\kappa \Lambda} \leq -\frac{4}{3}.
\end{align}
If this condition is satisfied, \textit{e.g.} by choosing $\Lambda$ or $\alpha$ appropriately, \textit{any} stable Yang--Mills solution will yield a sensible cosmological solution, those being oscillations around one of the minima of $W(a)$.

\section{\bf The case of general warping}\label{sec4:warping}
Up until now, we have only considered trivial products of $\mathbbm{R}\times G/H$, that is, the metric did not include any warping function. The simplicity of the CSDR construction makes it straightforward to generalize to the warped case, which is the topic of the current section. A particular choice of warping was used in \cite{Lechtenfeld_2018} to express the de Sitter space dS$_{n+1}$ via the usual spacelike $S^n$ slicing, thus obtaining gauge fields on dS$_{n+1}$ from the $\mathbbm{R}\times S^n$ CSDR construction. In that case, the warping resulted in a time-dependent friction term for $\phi(t)$. We will see that this idea generalizes for arbitrary warping, resulting in Hubble-friction-like terms. At the end, we will consider a particular choice of warping to obtain dynamics on the anti-de Sitter space AdS$_{n+1}$ from the well-known hyperbolic slicing $\mathbbm{R}\times H^n$, alluding to the aforementioned `dual' case of \cite{Lechtenfeld_2018}.

\subsection{Warpings as conformal transformations}

The metric of a warped cylinder $\mathbbm{R}\times_{a}\tilde{M}$ with warping function $a(u)$ is given by
\begin{align}
g=\de u\otimes\de u - a^2(u)g_{\tilde{M}}\,.
\end{align}
As in the FLRW case, these can always be made conformally flat with the introduction of a conformal `time' $\tau$ via $\de\tau = \frac{1}{a}\de u$ such that 
\begin{align}
g=a^2(\tau)(\de\tau\otimes\de\tau -g_{\tilde{M}})\,.
\end{align}
Therefore, instead of rescaling hypersurfaces to consider warped cylinders, we can always directly work with the equivalent conformal rescaling of a flat cylinder\footnote{That is, our foliation parameter $u\in\mathbbm{R}$ is set to be the conformal `time'.}. Now, let $g$ be the flat cylinder metric of the CSDR construction \eqref{metric}. Under a conformal transformation
\begin{align}
g\mapsto e^{2\sigma(u)}g\,, \quad \text{with}\quad e^{\sigma(u)} = a(u)\,,
\end{align}
the reduced CSDR action \eqref{reduced action} transforms as
\begin{align}
S[\phi]=\mathrm{Vol}(G/H)\int_{\mathbbm{R}} \mathcal{L}\,\de u \mapsto \mathrm{Vol}(G/H)\int_{\mathbbm{R}}e^{(n-3)\sigma(u)} \mathcal{L} \,\de u =:S^{(\sigma)}[\phi]\,.
\end{align}
Introducing the `conformal Hubble parameter' $\mathcal{H}(u)$ as 
\begin{align}
\mathcal{H}(u):=e^{-\sigma(u)}\frac{\de}{\de u}e^{\sigma(u)}=\dot{\sigma}(u)=\frac{\dot{a}(u)}{a(u)}\,,
\end{align}
the equations of motion for the warped case become
\begin{align}
\frac{\del\mathcal{L}}{\del\phi}-\frac{\de}{\de u}\frac{\del\mathcal{L}}{\del\dot{\phi}}-(n-3)\mathcal{H}(u)\frac{\del\mathcal{L}}{\del\dot{\phi}}=0\,.
\end{align}
Notice that the conformal invariance of the Yang--Mills theory in four spacetime dimensions is captured in the $(n-3)$ factor. Hence, the dynamics of the gauge field only changes by the addition of a Hubble-friction term. For the systems discussed in Sec.~\ref{sec:ApplyingCSDR}, including SO$(n+1)$/SO$(n)\cong S^n$, the equation of motion then reads
\begin{align}\label{hubble fric}
\ddot{\phi}+V'(\phi)+(n-3)\mathcal{H}(u)\dot{\phi}=0\,.
\end{align}
It is thus possible to generate equations of motion for $G$-invariant Yang--Mills fields for any spacetime conformally equivalent to $\mathbbm{R}\times G/H$, where $G/H$ can be $H^n$, $S^n$, dS$_n$ or AdS$_n$. They always reduce to a single scalar-like degree of freedom subjected to a double well or inverted double well potential together with the corresponding Hubble-friction term. However, as mentioned in Sec.~\ref{sec3:GRA}, notice that, outside of four spacetime dimensions, the equations of motion for the gauge and the gravity subsystems do not decouple anymore and the equation of state becomes substantially more complicated. In those cases, analytic solutions are in general not available, and more involved analyses are needed to probe the dynamics of the universe.

\subsection{Hyperbolic slicing of \texorpdfstring{AdS$_n$}{AdSn}}

To illustrate the previous discussion, we now briefly treat a particular example of warping with the hyperbolic slicing of the anti-de Sitter space AdS$_n$. As previously mentioned, a `dual' case was discussed in \cite{Lechtenfeld_2018}, where the de Sitter space $dS_{n+1}$ was foliated with spheres $S^{n}$ via the warping
\begin{align}
g_{\text{dS}_{n+1}}=\de t\otimes\de t -\cosh^2 t \, g_{S^n} = \frac{1}{\cos^2u} (\de u\otimes\de u -g_{S^n})\,.
\end{align}
For the simplest case $S^n \cong$ SO$(n+1)$/SO$(n)$, where only one degree of freedom remains, the resulting equation of motion is
\begin{align}
\ddot{\phi} \underbrace{+\frac{1}{2}\mathcal{D}_{n}(\phi^2 -1)\phi}_{V'_{S^n}(\phi)}+(n-3)\tan u \,\dot{\phi} =0\,,
\end{align}
which is reproduced by the general form \eqref{hubble fric}. In the same fashion, we can now slice AdS$_{n+1}$ with $H^{n}$ using the usual warping
\begin{align}
g_{\text{AdS}_{n+1}}=\de t\otimes \de t - \cos^2 t \, g_{H^n} = \frac{1}{\cosh^2 u}(\de u \otimes\de u - g_{H^n})\,.
\end{align}
The equation of motion then becomes
\begin{align}\label{eq:gauge-eom-hyperbolic}
\ddot{\phi}\underbrace{-\frac{1}{2}\mathcal{D}_{n}(\phi^2 -1)\phi}_{V'_{H^n}(\phi)} -(n-3)\tanh u \,\dot{\phi} =0\,,
\end{align}
whose solutions yield SO$(1,n)$ invariant gauge fields on AdS$_{n+1}$.

\begin{figure}[hbt!]
\centering
\includegraphics[scale=0.45]{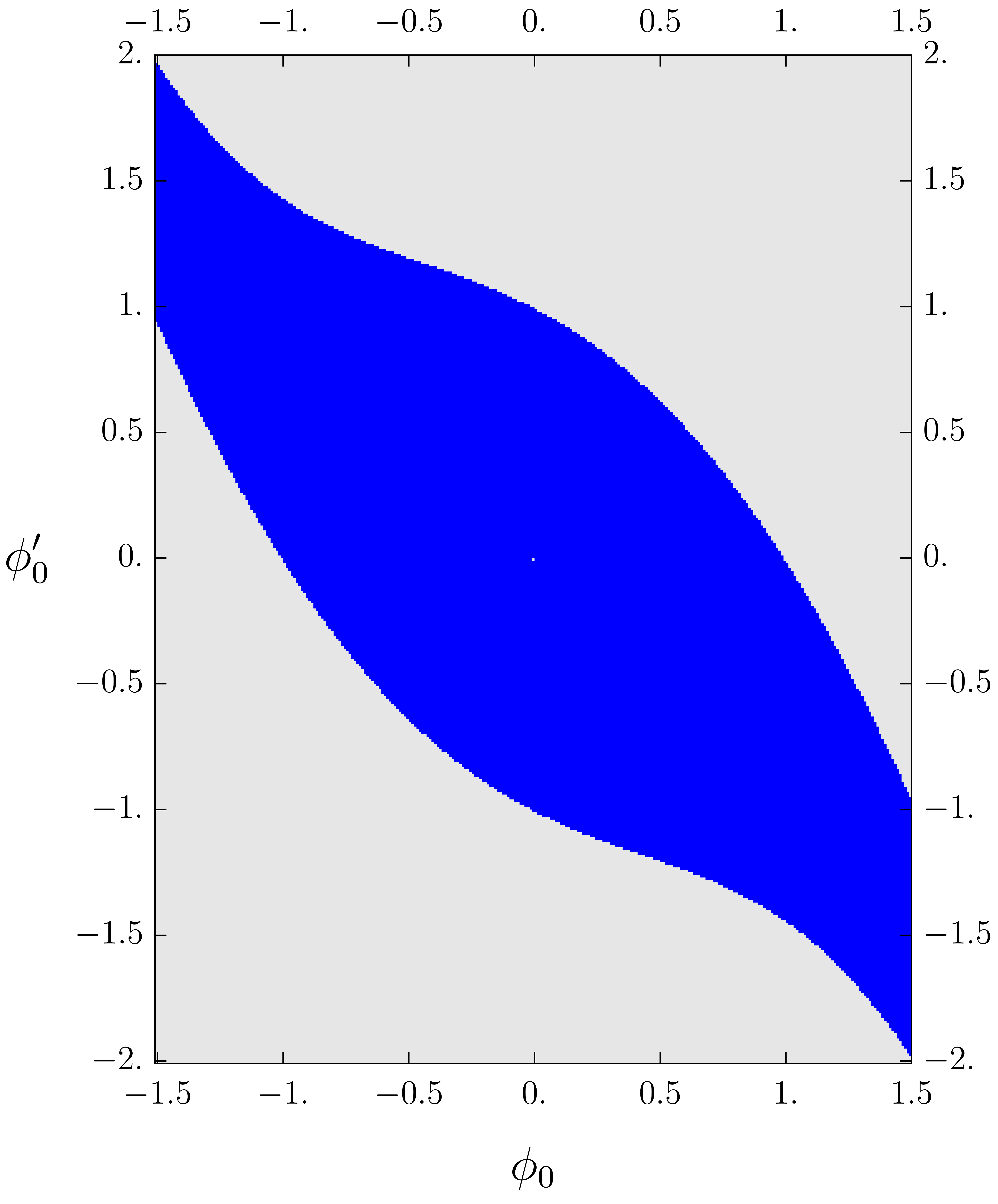}
\caption{Regions in phase space of initial conditions resulting in bounded (blue) or unbounded (gray) trajectories for the AdS$_3\cong \mathbbm{R}\times H^2$ warped case and initial (conformal) time $u_0 = 0$.}
\label{fig:initcond}
\end{figure}

Naturally, there are no analytic solutions available when $n\neq 3$ and a precise analysis involving the dynamics of the scale factor would require solving the full system, including the Friedmann equations. However, we can still extract some qualitative features of the gauge dynamics from \eqref{eq:gauge-eom-hyperbolic}, in a limit where the gauge field has a negligible impact on the scale factor dynamics. For the spherical slicings of dS$_{n+1}$ \cite{Lechtenfeld_2018}, the friction term is dissipative for $n>3$, making any solution freeze into one of the extrema of the double well, whereas for $n<3$ the `negative friction' term indicates an initial quick increase of $\phi$ and that our approximation ceases to be valid, with the scale factor dynamics potentially slowing down the acceleration of $\phi$ in the full system. In the AdS$_{n+1}$ case the situation is reversed, as the sign of the friction term is flipped. Also, in this case $\phi$ is subject to an inverted double well instead of an usual one. For $n>3$, despite the fact that the conformal Hubble parameter stays bounded, the inverted double well causes practically any initial condition, besides the field resting in one of the potential extrema, to very quickly increase, as the negative friction works to push $\phi$ out of the well. On the other hand, for $n=2$, the friction enlarges the set of initial conditions for which $\phi$ stays bounded. This can be understood intuitively, since, with an adequate tuning of initial position and velocity and provided that the approximation on the vacuum scale factor is still valid, $\phi$ can start with initial energy larger than the height of the inverted double well in such a way that $\phi$ will find itself inside the well after enough dissipation, when its energy reaches the height of the well. Consequently, it would end up in the origin of the phase space after enough time\footnote{evidently, there is also a set of measure zero of initial conditions such that the particle ends up exactly at one of the two maxima of the potential, which is the boundary of region of initial conditions with bounded solutions.}. The region of initial conditions resulting in bounded solutions depends on the initial (conformal) time $u_0$ and can be evaluated numerically, as illustrated in the Fig.~\ref{fig:initcond} for $u_0 = 0$.

\section{\bf Conclusion}\label{sec5:conclusion}
\setlength{\parskip}{\baselineskip}

In previous works using geometric methods to construct non-trivial Yang--Mills fields, the reduction of the dynamics from a system of non-linear partial differential matrix equations to a single `scalar' ordinary differential equation was already seen for particular cases, where the full gauge dynamics emerge from that of a Newtonian particle subjected to a (inverted or not) quartic potential. In particular cases, the coupling to gravity was first considered aiming applications in cosmology, but mostly regarding the four-dimensional closed FLRW universe foliated by three-spheres.

In this paper, we used the coset space dimensional reduction scheme to show that a construction that highly simplifies the gauge dynamics is always possible for a variety of Lorentzian spacetimes given by (warped or not) cylinders over symmetric spaces, of which $S^n$, $H^n$, dS$_n$, and AdS$_n$ are especially relevant for Lorentzian spacetimes. Moreover, we discuss that coupling to gravity through the cosmological scale factor, obtaining the reduced equations of motion for the fully coupled system in any spacetime dimension and discussing a couple of particular cases of interest, including checks against previous results in the literature for the most studied scenario of dS$_4 \simeq \mathbbm{R}\times S^3$. A particularly interesting result we derived is that, for any spacetime dimension, it is always possible to choose a time coordinate in any warped foliations such that the reduced equation of motion for the gauge sector only picks up a Hubble friction-like term when compared to the conformally invariant scenario of four-dimensional spacetimes, in which the dynamics of the gauge and gravity sectors decouple.

As we have seen throughout this paper, the CSDR approach is a very useful tool to find highly treatable but still non-trivial dynamics for the full Einstein--Yang--Mills system. This work sets the stage for future research directions which fully explore such systems in FLRW universes, considering possible cosmological applications. Investigations could entail three or higher dimensional general relativity, with the possibility of the dynamics described by the Hubble friction-like term when $n\neq 3$ to play a role in dark matter or dark energy physics, or, in four-dimensional spacetimes, the reduced dynamics of highly symmetric non-Abelian gauge fields coupled to an inflaton or to axion-like particles, or with the addition of the gauge $F^4$ interaction term necessary for traditional gauge-flation scenarios.

\section*{\bf Acknowledgements}

ME wants to thank Prof.\@ Olaf Lechtenfeld and Prof.\@ Domenico Giulini for many important discussions.

\bibliographystyle{JHEP}
\bibliography{references,Bib-PhD-benty,Bib-PhD-non-benty}

\end{document}